\documentstyle[12pt]{article}
\catcode`\@=11
\begin{document} \openup6pt

\title{PROBABILITY FOR  PRIMORDIAL BLACK HOLES IN HIGHER DERIVATIVE THEORIES
}

\author{B. C. Paul\thanks{ e-mail : bcpaul@nbu.ernet.in }   \\
   Department of Physics, North Bengal University, \\
   Siliguri, Dist. Darjeeling, Pin : 734 430, INDIA \\
Arindam Saha  \\
Lataguri H. S. School, 
P.O. : Lataguri \\ Dist. : Jalpaiguri, West Bengal, Pin 735 219, INDIA }

\date{}

\maketitle

\begin{abstract}

The probability for quantum creation of an inflationary universe with a
 pair of black holes in higher derivative theories has been studied. 
Considering a gravitational action which includes quadratic 
($\alpha R^{2}$) and/or cubic term ($\beta R^{3}$) in scalar
 curvature in addition to a cosmological constant ($\Lambda$) in 
semiclassical approximation with Hartle-Hawking boundary condition, 
the probability has been evaluated. The action of the instanton responsible 
for creating such a universe, with spatial section with
 $S^{1}XS^{2}$ topology, is found to be less than that with a spatial 
 $S^{3}$  topology, unless $\alpha < - \frac{1}{8 \Lambda}$ in 
$R^{2}$-theory. In the $R^{3}$ theory, however, there exists a set 
of solutions without a cosmological constant when $\beta R^{2} = 1$ and
 $\alpha = - 3 \sqrt{\beta}$ which admit primordial black holes (PBH) 
pair in an inflationary universe scenario. We note further that when 
$\beta R^{2} \neq 1$, one gets PBH pairs in the two cases : 
(i) with $\alpha$ and $\Lambda$ both positive and (ii) with 
$\Lambda$ positive and $\alpha$ negative satisfying a constraint 
$6 | \alpha | \Lambda > 1$. However, the relative probability for 
creation of an inflationary universe with a pair of black holes in the 
$R^{3}$-theory suppresses when $\alpha > -  2 \sqrt{\beta} $ or 
$|\alpha| <  2 \sqrt{\beta} $. However, if the above constraints are 
relaxed one derives interesting results leading to a universe
with PBH in $R^{3}$-theory without cosmological constant. \\
PACS No(s). : 04.20.Jb, 04.60.+n, 98.80.Hw
\end{abstract}

\pagebreak

{\bf I. INTRODUCTION : }

In recent times black holes are the most elusive object to study  in
 Astrophysics and Cosmology. The mass of these objects may be greater than
 the solar mass or even less. It is known in stellar physics that a 
blackhole is the ultimate corps of a collapsing star when its mass exceeds 
twice the  mass of the sun. Another kind of black holes are becoming 
important in cosmology which might have formed due to quantum fluctuations 
of matter distribution in the early universe. These are topological
 blackholes having mass very small.  In particular these are important 
from the view point of Hawking radiation [1]. Bousso and Hawking [2] 
( in short, BH ) calculated the probability of the quantum creation of 
a universe 
with a pair of primordial black holes in  (3 + 1) dimensional universe. 
In the paper they considered two different 
euclidean space-time : (1) a universe with space-like sections with 
topology $S^{3}$ and (2) a universe with space-like section with topology 
$ S^{1} \times S^{2}$ , as is obtained in the Schwarzschild- de Sitter
 solution.
The first kind of spatial structure describes an 
inflationary (de Sitter) universe without  black hole while the second kind
 describes a Nariai 
universe [3], an inflationary universe with a pair of black holes.  
BH considered in their paper a massive scalar field which provided an 
effective cosmological constant for a while through a slow-rolling potential 
(mass-term).
 Chao [4] has studied the creation of a primordial black hole and
 Green and Malik [5] studied the primordial black holes production during
 reheating. Paul {\it et al.} [6] following the approach of BH studied the
 probability
of creation of PBH including $R^{2}$-term in the Einstein action and found
that the probability is very much suppressed in the $R^{2}$ -theory. 
It is well known that the theory with higher order Lagrangian are 
conformally equivallent to Einstein gravity with a matter sector 
containing a minimally coupled, self interacting scalar field. 
However, the renormalization of higher loop contributions 
introduces terms into the effective action that are higher than 
quadratic in $R$. Consequently it is important to study the effects of 
these terms to study quantum creation of a universe with a pair of PBH.  
It is important to confine with terms upto cubic terms only as for a
 polynomial Lagrangian
\[
f(R) = \sum_{n=1}^{N} \lambda_{n} R^{n},
\]
 the form of the potential is extremely complicated. It  was shown by 
Henk van Elst {\it et al.} [7] that $N = 2$ contribution is rather special 
in four dimensions, while  $R^{3}$-term gives a more generic purturbation.
 The qualitative behavior of the potential $V(\phi)$ does not change
 relative to the $R^{3}$ term even when terms with $N > 3$ are considered. 
We, therefore, consider term upto $R^{3}$ in the action and look for 
probability of a pair of PBH in the early universe. 
We calculate the probabilities for the creation  of a universe with two
 types of topology namely, $ R \times S^{3} $ topology and 
$R \times S^{1} \times S^{2}$ topology  which accommodates
a pair of primordial black holes. 
To calculate the probabilities for these spatial topologies,   we use a 
semiclassical approximation for the evaluation of  the 
euclidean path integrals. The condition that a classical spacetime 
should emerge, to a good 
approximation, at a large Lorentzian time was selected by a choice of the
 path of the time parameter $\tau$ along the $\tau^{Re}$ axis 
from 0 to $\frac{\pi}{2H}$ and 
 then continues along the $\tau^{im}$ axis. 
The  wave-function of the universe in the semiclassical approximation  is 
given by
\begin{equation}
\Psi_{o} [ h_{ij}  , \Phi_{\partial  M} ]   \approx \sum_{n} A_{n}
 e^{- I_{n}}
\end{equation}
where the sum is over the saddle points of the path integral, 
and $I_{n}$
denotes the corresponding Euclidean action.  The probability measure of the
creation of PBH is
\[
P[ h_{ab} , \Phi_{\partial M} ] \sim  e^{ - 2 \;  I^{Re}}
\]
where $h_{ab}$ is the boundary metric and  $I^{Re}$  is the real part of
the action corresponding to the dominant saddle point, i.e. the classical
solution satisfying the Hartle-Hawking (henceforth, HH) boundary
conditions [8].  It was believed that all inflationary models lead
to $ \Omega_{o} \sim 1$ to a great accuracy. This view was modified after it
was discovered  that there is a special class of inflaton effective 
potentials which
may lead to a nearly homogeneous open universe with  $ \Omega_{o} \leq 1$ 
at the present epoch. Cornish {\it et al.} [9, 10] studied the problem of 
pre-inflationary
homogeneity and outlines the possibility of creation of a small, compact,
 negatively
curved universe. We show that a universe with $S^{3}$-topology may give 
birth to
an open inflation.

The paper is organised as follows : in sec. II  we  write the gravitational
 action for
a higher derivative theories and obtain gravitational instanton solutions 
and
in sec. III we use the action to estimate the relative probability of the 
two types of the universes and in sec. IV we give a brief discussion.

\vspace{1in}

{\bf II. GRAVITATIONAL INSTANTON SOLUTIONS   
WITH OR \\ 
WITHOUT A PAIR OF PRIMORDIAL BLACK HOLES: }

We consider a  Euclidean action  containing higher derivative terms which is
given by
\begin{equation}
I_{E} =  - \frac{1}{16\pi} \int d^{4} x \sqrt{g}  f(R)  - 
\frac{1}{8\pi} \int_{\partial M} d^{3} x \sqrt{h} K f'(R) 
\end{equation}
where $g$ is the 4-dimensional Euclidean metric, $f(R) = R + \alpha R^{2} + 
\beta R^{3} - 2 \Lambda $, $R $ is the Ricci scalar
 and $\Lambda$ is the 4-dimensional cosmological constant.
In the gravitational surface term, $h_{ab}$ is the boundary metric and 
$K = h^{ab} K_{ab}$ is the trace of the second fundamental form of the
 boundary
$\partial M$ in the metric. The second term is the contribution from
 $\tau = 0$ back in the action. It vanishes for a universe with $S^{3}$ 
topology, but gives a non vanishing contribution for 
$S^{1} \times S^{2}$-topology.

{\bf (A)  Topology $S^{3}$, the de Sitter spacetime :}

In this section we study vacuum solutions of the Euclidean Einstein equation 
with a cosmological constant in four dimensions. We now look for a solution 
with spacelike section $S^{3}$. Hence we choose the
four dimensional metric ansatz
\begin{equation}
ds^{2} = d\tau^{2} + a^{2} (\tau) \left[ dx_{1}^{2} + sin^{2} x_{1} 
d\Omega_{2}^{2} \right]
\end{equation}
where $a(\tau)$ is the scale factor of a four dimensional universe and
 $ d\Omega_{2}^{2} $ is a line element on the unit two sphere.
The scalar curvature is given by
\[
R = -  6 \left[ \frac{\ddot{a}}{a} + \left(
    \frac{\dot{a}^{2}}{a^{2}} - \frac{1}{a^{2}} \right) \right] .
\]
where an overdot denotes differentiation with respect to $\tau$. We rewrite 
the action (2), including the constraint through a Lagrangian multiplier 
$\beta$ and obtain 
\begin{equation}
I_{E} = -  \frac{\pi}{8}  \int 
         \left[ f(R) a^{3} - \beta \left( R + 6 \frac{\ddot{a}}{a} + 
6 \frac{\dot{a}^{2} -1}{a^{2}} \right) \right] d \tau 
- \frac{1}{8 \pi} \int_{\partial M} d^{3} x \sqrt{h} K f'(R).      
\end{equation}
Varying the action w.r.t. R, we determine
\begin{equation}
\beta = a^{3} f'(R).
\end{equation}
Substituting the above equation, treating $a$ and $R$ as independent 
variables we get 
\[
I_{E} = -  \frac{\pi}{8}  \int_{\tau = 0}^{\tau_{\frac{\pi}{2 H_{o}}} } 
         \left[ a^{3} f(R) - f'(R) \left( a^{3} R - 6  a \dot{a}^{2} - 6 a 
\right)  + 6 a^{2} \dot{a} \dot{R} f''(R)  \right] d \tau
\]
\begin{equation}
-  \frac{3 \pi}{4} \left[  \dot{a} a^{2} f'(R) \right]_{\tau = 0},
\end{equation}
here we have eliminated $\ddot{a}$ by integration by parts. The field 
equations can now be obtained by varying $I_{E}$ with respect to $a$  and
 $R$ , 
giving
\begin{equation}
f''(R) \left[ R + 6  \frac{ \ddot{a}}{a}    + 
6 \frac{\dot{a}^{2} - 1}{a^{2}} \right]  = 0 ,
\end{equation}
\begin{equation}
2 f'''(R) \dot{R}^{2} + 2 f''(R) \left[ \ddot{R} + 2 \frac{\dot{a}}{a} 
\dot{R} \right] + f'(R) \left[ 4 \frac{ \ddot{a}}{a} + 2 
\frac{ \dot{a}^{2}}{a^{2}} - \frac{2}{a^{2}} + R \right] - f(R)  = 0.
\end{equation}
We now consider $f(R) = R + \alpha R^{2} + \beta R^{3} - 2 \Lambda $ and
 obtain  an instanton solution which is given by
\begin{equation}
a = \frac{1}{H_{o}} \; sin  H_{o} \tau 
\end{equation}
where
$ H_{o}^{2} [ \beta, \Lambda ] $ is determined from the constraint equation 
$ 432 \beta H_{o}^{6} - 3 H_{o}^{2} + \Lambda = 0 $.
We note that this solution satisfies the HH no boundary conditions viz., 
$a(0) = 0 $, $ \dot{a} (0) = 1$. One can choose a path along the $\tau^{Re}$
axis to  $\tau = \frac{\pi}{2 H}$, the solution describes half of the 
Euclidean
de Sitter instanton $S^{3}$.  Analytic continuation of the metric (3) to 
Lorentzian region, $x_{1} \rightarrow \frac{\pi}{2} + i \sigma $, gives
\begin{equation}
ds^{2} = d \tau^{2} + a^{2}(\tau) \left[ - d\sigma^{2} + \cosh^{2} \sigma
 d \Omega_{2}^{2} \right]
\end{equation}
which is a spatially inhomogeneous de Sitter like metric. However, if one 
sets $\tau = i t $ and  $ \sigma = i \frac{\pi}{2} + \chi $, the metric 
becomes
\begin{equation}
ds^{2} = -  dt^{2} + b^{2}(t) [ d\chi^{2} + \sinh^{2} \chi
 d \Omega_{2}^{2} ]
\end{equation}
where $b(t) = - i a( it )$. It describes the creation of an open 
inflationary  universe.
The real part of the Euclidean action 
corresponding to the solution calculated by following the complex contour
of $\tau $ suggested by BH is given by
\begin{equation}
I^{Re}_{S^{3}} = - \frac{\pi}{2 H_{o}^{4}} \left[ 24 \alpha H_{o}^{4} + 
4 H_{o}^{2} - \Lambda \right].
\end{equation}
With the chosen path for $\tau $ ,  the solution describes 
half the de Sitter Instanton  with $S^{4}$ 
topology, joined to a real Lorentzian hyperboloid of
topology  $R^{1} \times S^{3}$. It can be joined to any boundary satisfying 
the condition 
$a_{\partial M} > 0$ . 
For $a_{\partial M} > H_{o}^{- 1}$ , the wave function oscillates and 
predicts a classical
space-time. 
We note the following cases :
$\bullet$ $\beta = 0$ one obtains $H_{o}^{2} = \Lambda/3$ 
and $R = 4 \Lambda$ which corresponds to the result obtained by
 Paul {\it et al.} [6].
$\bullet$ $\beta \neq 0$ and one gets now 
instanton solution even with a cosmological constant i.e., 
$\Lambda = 0$ giving $H_{o}^{2} = \frac{1}{12 \sqrt{\beta}}$. 
This solution is interesting as with $N \leq 2$ in the
polynomial $f(R)$ one always requires a cosmological constant to get such 
an instanton. 

{\bf (B) Topology $ S^{1} \times S^{2} $ :}

In this section we consider  Euclidean Einstein equation
 and look for a universe with $S^{1} \times
S^{2}$ -spacelike sections as this topology accommodates a pair of black 
holes. The corresponding ansatz for (1 + 1 + 2) dimensions is given by
\begin{equation}
ds^{2} = d \tau^{2} + a^{2}(\tau) dx^{2} + b^{2}(\tau) \;  d \Omega_{2}^{2} 
\end{equation}
where $a( \tau ) $ is the scale factor of two sphere and $b( \tau )$ is the
scale factor of the two sphere given by the metric 
\[
d\Omega_{2}^{2} = d \theta^{2} + sin^{2} \theta \; d \phi^{2} 
\]
The scalar curvature is given by
\begin{equation}
R = - \left[ 2 \frac{\ddot{a}}{a} +  4 \frac{\ddot{b }}{b}
     + 2 \left( \frac{\dot{b}^{2}}{b^{2}} - 
   \frac{1}{b^{2}} \right) + 4 \frac{\dot{a} \dot{b}}{a b} \right] .
\end{equation}
The Euclidean action (2) becomes
\[
I_{E} = - \frac{\pi}{2} \int
 \left[  f(R) a b^{2} - \beta \left(R + 2 \frac{\ddot{a}}{a} + 4
\frac{\ddot{b}}{b} + 4 \frac{\dot{a} \dot{b}}{ab} + 2 
\frac{\dot{b}^{2}}{b^{2}} - \frac{2}{b^{2}}\right) \right] d\tau
\]
\begin{equation}
 - \frac{1}{8 \pi} \int_{\partial M} \sqrt{h} d^{3}x K f'(R).
\end{equation}
One can determine $\beta$ as is done before and obtains 
\[
I_{S^{1} \times S^{2}} = - \frac{\pi}{2} 
\int_{\tau = 0}^{\tau_{\partial M}} 
 \left[  f(R) - f'(R) \left( R - 4  \frac{\dot{a} \dot{b}}{ab} - 
2 \frac{\dot{b}^{2}}{b^{2}} - \frac{2}{b^{2}}\right) + 2 f''(R)  
\dot{R} \left( \frac{\dot{a}}{a} + 2 \frac{\dot{b}}{b} \right) \right] 
\]
\begin{equation}
a b^{2} d \tau - \pi \left[ \left( \dot{a} b^{2}  + 2 a b \dot{b} \right) f'(R) 
\right]_{\tau = 0}.
\end{equation}
Variation of the action with respect to the scale factors $R $, $a$ and $b$
 gives
the following field equations
\begin{equation}
f''(R) \left( R + 2 \frac{\ddot{a}}{a}
  + 4 \frac{\ddot{b}}{b} + 4 \frac{\dot{a} \dot{b}}{ab}  
+ 2 \frac{\dot{b}^{2}}{b^{2}} - \frac{2}{b^{2}} \right) ,
\end{equation}
\begin{equation}
2  f'''(R) \dot{R}^{2} + 2 f''(R) \left( \ddot{R} + 2 \dot{R} 
\frac{\dot{b}}{b}
\right) + f'(R) \left( R + 4 
\frac{\ddot{b}}{b}  +  
2 \frac{ \dot{b}^{2} - 1}{b^{2}} \right) - f(R)  = 0,
\end{equation}
\[
2 f'''(R) \dot{R}^{2} + 2 f''(R) \left( \dot{R} \left(\frac{\dot{a}}{a} +
\frac{\dot{b}}{b}\right)  + \ddot{R} \right) + f'(R)  
\left( R + 2  \frac{\ddot{a}}{a} +
 2 \frac{\ddot{b}}{b}  + 2 \frac{\dot{a} \dot{b}}{ab} \right) 
\]
\begin{equation}
- f(R)   = 0.
\end{equation}
With $f(R)  = R + \alpha R^{2} + \beta R^{3} - 2 \Lambda $ one obtains   
an instanton solution from the field equations  (17)-(19) which is given by
\[
a = \frac{1}{H_{o}} \; sin ( H_{o} \tau ) , \; \; \;
b =  H_{o}^{- 1},
\]
\begin{equation}
R = 4 H_{o}^{2} 
\end{equation}
where $H_{o}$ satisfies the equation 
\[
16 \beta H_{o}^{6} - H_{o}^{2} + \Lambda = 0.
\]
 We note the following : (i) when $ \Lambda = 0$, $H_{o}^{4} = 
\frac{1}{16 \beta}$ and $\alpha= - 3 \sqrt{\beta}$ which demands a 
positive coupling constant $\beta$, (ii) when 
$ \Lambda \neq  0$ then $\alpha= \pm 3 \sqrt{\beta}$ which demands
 a positive coupling constant $\beta$ but $\alpha$ may take either sign, 
(iii) when $\beta = 0$, $\Lambda = H_{o}^{2}$ corresponds to the solution 
obtained by Paul {\it et al.} [6], and (iv) $\alpha = 0$ and $\beta = 0$ 
leads to $\Lambda = H_{o}^{2}$ which corresponds to BH solution [2].
These solutions satisfy the HH boundary conditions
$a(0) = 0 $, $ \dot{a} (0) = 1,  b(0) = b_{o} $, $ \dot{b} (0) = 0$.
Analytic continuation of the metric (13) to Lorentzian region, i.e.,
$\tau \rightarrow it $ and $ x \rightarrow \frac{\pi}{2} + i \sigma $ yields
\begin{equation}
ds^{2} = -  dt^{2} + c^{2}(t) d\sigma^{2} + H_{o}^{-2} d \Omega_{2}^{2}
\end{equation}
where $c(t) = - i a( it )$. In this case the analytic continuation of time 
and space
do not give an open inflationary universe. The corresponding Lorentzian 
solution
is given by
\[
a(\tau^{Im}) |_{\tau^{Re} = \frac{\pi}{2H}} = H^{-1} \cosh H \tau^{Im} ,
\]
\[
b(\tau^{Im}) |_{\tau^{Re} = \frac{\pi}{2H}} = H^{-1} 
\]
Its space like sections can be visualised as  three spheres of radius $a$
with a {\it hole} of radius $b$ punched through the north and south poles. 
The
physical interpretation of the solution is that of two - spheres containing
two black holes at opposite ends. The black holes have the radius $H^{-1}$
accelerates away from each other with the expansion of the universe.
The
real part of the action can now be determined following the contour
 suggested  by BH [2], and it is given by
\begin{equation}
I^{Re}_{S^{1} X S^{2}} = - \frac{\pi}{ H_{o}^{4}} \left[ 8 \alpha  H_{o}^{4}
+ 4 H_{o}^{2} - 3 \Lambda \right].
\end{equation}
where $H_{o}^{2}$ satisfies the constraint 
\[
16 \beta H_{o}^{6} - H_{o}^{2} + \Lambda = 0.
\]
The solution (20) describes a universe with two black holes at the poles
of a two sphere. However, unlike the BH case we are  getting 
vanishing contribution of the action in this case  from the surface term.

{\bf III. EVALUATION FOR THE PROBABILITY FOR PBH :}

In the previous section we have calculated the actions for inflationary 
universe with or without a pair of black holes. We now compare the
 probability
measure.
The probability for creation of a higher dimensional de Sitter universe 
may be obtained from the action (6).  Thus, the probability for
nucleation of a higher dimensional universe
without PBH is given by
\begin{equation}
P_{S^{3}} \sim \; e^{ \frac{\pi}{H_{o}^{4}} \left[ 24 \alpha H_{o}^{4} +
 4 H_{o}^{2} -  \Lambda \right] }.
\end{equation}
However for an inflationary universe with a pair of black holes the 
corresponding probability of nucleation can be obtained from the
action (16).  The corresponding probability is
\begin{equation}
P_{S^{1} \times S^{2}} \sim e^{ \frac{2 \pi}{H_{o}^{4}} \left[  
 8 \alpha H_{o}^{4} + 4 H_{o}^{2} - 3    \Lambda \right] }
\end{equation}
We now describe the spacial cases of the general result :

$\bullet$ when $\alpha = 0$, $\beta = 0$ one  recovers the result obtained by 
Bousso and Hawking
\begin{equation}
P_{S^{3}} \sim e^{\frac{3 \pi}{\Lambda}} , \; \; \; P_{S^{1} 
\times S^{2}} \sim e^{\frac{2 \pi}{\Lambda}}.
\end{equation}
Thus with a positive cosmological constant the probability for a
 universe with PBH is less than that without PBH.

$\bullet$ when $\alpha \neq  0$, $\beta = 0$ one  determines 
$H_{o}^{2} = \Lambda$ and the probabilities are 
\begin{equation}
P_{S^{3}} \sim e^{\frac{3 \pi}{\Lambda} + 
24 \pi \alpha},  \; \; \; P_{S^{1} \times S^{2}} 
\sim e^{\frac{2 \pi}{\Lambda} + 16 \pi \alpha}.
\end{equation}
In this case with a positive cosmological constant and a positive 
$\alpha > 0$, 
de Sitter universe is more probable [6]. 
\begin{equation}
P_{S^{3}} \sim e^{\frac{3 \pi}{\Lambda}},  \; \; \; P_{S^{1} 
\times S^{2}} \sim e^{\frac{2 \pi}{\Lambda}}.
\end{equation}
However negative values of $\alpha < - \frac{1}{8 \Lambda}$ 
could lead to an interesting possibilities as the probability for 
a universe with PBH in this case is more.However, the case 
$\alpha < 0$ leads to a classical instability in $R^{2}$-theory.

$\bullet$ when $\alpha \neq  0$, $\beta \neq 0$ one  determines 
obtains instanton solution even with $ \Lambda = 0$ and we get 
$H_{o}^{2} = \sqrt{\frac{1}{144 \beta}}$ 
which leads to  the probabilities  
\begin{equation}
P_{S^{3}} \sim e^{ 8 \pi \left( 3 \alpha + 6 \sqrt{\beta} 
\right)}, \; \; \; P_{S^{1} \times S^{2}} \sim e^{ 16 \pi 
\left( \alpha + 2 \sqrt{\beta} \right)} .
\end{equation}
It is evident that one requires 
$\beta > 0$ to obtain instanton but the values of
 $\alpha$ may be positive or negative.  When $ \alpha > - 2 \sqrt{\beta}$ 
or $|\alpha| <  2 \sqrt{\beta}$ one obtains that a universe without PBH is 
more probable.
In this case interesting possibilities emerges when $ \alpha 
< -  2 \sqrt{\beta}$ or $|\alpha|  >  2 \sqrt{\beta}$.

{\bf IV. DISCUSSIONS : }

In this work we have evaluated the probability for primordial
black holes pair creation in a higher derivative theories. In
section II, we have obtained the gravitational instanton solutions in two 
cases : a universe with (i) $R \times S^{3}$ - topology  and a universe with
(ii) $ R \times S^{1} \times S^{2} $ -topology respectively. The 
Euclidean  action is then evaluated. We found  that the probability of a
 universe
with $R \times S^{3} $ topology turns out to be much lower than a universe
 universe with topology $ R \times S^{1} \times S^{2} $ 
in the $R^{2}$-theory unless $\alpha < - \frac{1}{8 \alpha}$.
It may be mentioned here that one gets a regular instanton in 
$S^{4} $ - topology
if there are no black holes. The existence of black holes  
restricts such a regular topology. 
The result obtained here on the probability of creation of a universe with a  pair of primordial black holes is found to be strongly 
suppressed if one extends the theory with  $R^{3}$-term in the action under
the constraints (i)  $ \alpha > -  2 \sqrt{\beta}$ or (ii) $|\alpha| 
<  2 \sqrt{\beta}$ when $\Lambda = 0$. Thus one interesting result is that
 in the framework of $R^{3}$-theory we get de Sitter instantons with 
$S^{3}$ and $S^{1} \times S^{2}$ topologies even without a cosmological
 constant. 
Thus one obtains PBH in the theory even without a cosmological
constant in $R^{3}$ theory.
It is interesting to note here that analytic continuation of a 
$R \times S^{3}$ metric considered here
to Lorentzian region leads to an open 3 - space.  One
obtains Hawking-Turok [11, 12] type open inflationary universe in this case.
In the 
other type of topology  an open inflation is not permitted.
A detail study of an open inflationary universe will be discussed elsewhere.
Thus in a polynomial Lagrangian in $R$-theory, quantum creation of PBH seems 
to be suppressed in the minisuperspace cosmology for some values of the 
parameters in the action.

{\bf {\it Acknowledgement  :}}

The work is suported by Minor Research grant of the University Grants 
Commission, New Delhi and North Bengal University (NBU). BCP would like to 
thank the Inter-University University Centre for Astronomy
and Astrophysics (IUCAA), Pune for providing facilities to carry out the 
work and AS would like to thank the Physics Department, North Bengal 
University and 
to S. Mukherjee for providing IRC facilities at NBU to carry out a part 
of the work. 
\pagebreak


\begin{thebibliography}{99} 

\bibitem{kn : 1} S. W. Hawking, {\it Comm. Math. Phys.} {\bf 43}, 199 (1975).
\bibitem{kn : 2}  R. Bousso and S. W. Hawking, {\it Phys. Rev.} D {\bf 52},
5659 (1995). 
\bibitem{kn : 3}  H. Nariai,{\it Sci. Rep. Tohoku Univ.} {\bf 35}, 62 (1951).
 
\bibitem{kn : 4} W. Z. Chao, {\it Int. J. Mod. Phys.} {\bf D6}, 199 (1997).
\bibitem{kn : 5}A. M. Green and K. A. Malik, {\it  Phys. Rev. } D {\bf 64}, 
021301 (2001).
\bibitem{kn : 6}  B. C. Paul, G. P. Singh, A. Beesham and S. Mukherjee,
{\it Mod. Phys.
 Letts.}   {\bf A 13}, 2289 (1998). 
\bibitem{kn : 7}  Henk van Elst, J. E. Lidsey and R. Tavakol, {\it Class.
 Quantum Grav. } {\bf 11}, 2483 (1994). 
\bibitem{kn : 8}  J. B. Hartle and S. W. Hawking,{\it Phys. Rev.} 
D {\bf 28}, 2960 (1983).
\\bibitem{kn : 9}  N. J. Cornish, D. N. Spergel and G. D.  Starkman,
 {\it Phys. Rev. Letts.} 
 {\bf 77}, 215 (1996) ; {\it Class. Quantum Grav. } 
  {\bf 15}, 2657 (1998).
\bibitem{kn : 10}  N. J. Cornish and D. N. Spergel,  {\it ` A small Universe
after all ? ' } - astro-ph/9906401.
\bibitem{kn : 11}  S. W. Hawking and N. G. Turok, {\it  Open Inflation 
without 
false Vacuum}, hep-th/9802030.
\bibitem{kn : 12}  N. G. Turok and S. W. Hawking, {\it  Open Inflation,
the four form and the cosmological constant}, hep-th/9803156.
 
\end{thebibliography}
\end{document}